# Knowledge Distillation for Mobile Edge Computation Offloading

Haowei Chen, Liekang Zeng, Shuai Yu, and Xu Chen

*School of Data and Computer Science,*

*Sun Yat-sen University, Guangzhou, China*

**Abstract**

**Edge computation offloading allows mobile end devices to put execution of compute-intensive task on the edge servers. End devices can decide whether offload the tasks to edge servers, cloud servers or execute locally according to current network condition and devices' profile in an online manner. In this article, we propose an edge computation offloading framework based on Deep Imitation Learning (DIL) and Knowledge Distillation (KD), which assists end devices to quickly make fine-grained decisions to optimize the delay of computation tasks online. We formalize computation offloading problem into a multi-label classification problem. Training samples for our DIL model are generated in an offline manner. After model is trained, we leverage knowledge distillation to obtain a lightweight DIL model, by which we further reduce the model's inference delay. Numerical experiment shows that the offloading decisions made by our model outperforms those made by other related policies in latency metric. Also, our model has the shortest inference delay among all policies.**

**Keywords: mobile edge computation offloading, deep imitation learning, knowledge distillation.**

## 1 Introduction

Nowadays more and more end devices are running compute-intensive tasks, such as landmarks recognition apps in smartphones [1], vehicles detection apps used for counting traffic in cameras [2], augmented reality apps in Google Glass, etc. The advantages of executing compute-intensive tasks on end devices are twofold. On the one hand, most data, e.g., image, audio and video, are generated at end devices. Compared to sending these data to the cloud server, processing data locally on end devices can avoid time-consuming data transmission and reduce heavy bandwidth consumption. On the other hand, some tasks are sensitive to latency and the execution result can be out of date if being late. In some cases, e.g., face recognition applications, high latency can result in poor user experience. If computation tasks are offloaded to the cloud, the unreliable and delay-significant wide-area connection can be problematic. Hence, executing compute-intensive tasks on end devices is a potential solution to lower end-to-end latency.

However, compared with cloud servers, the computing resources of end devices is very limited. Even a smartphone's computing capability is far weaker than cloud server, not to mention the Google Glass and cameras. It turns out that executing compute-intensive tasks on end devices may result in high computation latency. In addition, end devices often have energy consumption restrictions, for example, most smartphone users do not want a single app to consume too much power. Thus, it is unwise to execute tasks on end devices indiscriminately.

Recently, edge computing has emerged as a new paradigm difference from local execution and cloud computing, and attracted more and more attention. The European Telecommunications Standards Institute provided a concept of multi-access edge computing (MEC) [3]. In the MEC architecture, distributed edge servers are located at the network edge to provide computing capabilities and IT services with high bandwidth and real-time processing. Edge servers become the third offloading location of compute-intensive tasks in addition to end devices and cloud. However, due to edge servers' restricted computing capability, they cannot completely take place of cloud servers. Many factors, including available computation and communication resources, should be taken into consideration when making offloading decisions. To tackle this challenge, in this paper, we design a computation offloading framework which jointly considers computation and communication and dynamically makes optimal offloading decisions to minimize the end-to-end execution latency.

Recent advances in deciding offloading strategies focuses on learning-based methods. Yu et al. [13] propose to "imitate" the optimal decisions of traditional methods by Deep Imitation Learning (DIL), where DIL [4] uses instances generated from human's behaviors to learn the decision strategies in specific environments. DIL enjoys two advantage compared to traditional methods (e.g. [7]) and deep reinforcement learning methods (e.g., [10]). First, inference delay of DIL is much shorter than traditional methods especially when the amount of input data is large (as shown in our experiment in Section 5). Second, DIL has higher accuracy in imitating optimal offloading decisions comparing to DRL-based approaches.

However, DIL model is built upon Deep Neural Network (DNN), which is compute-intensive and typically requires high inference latency. On this issue, model compression is proposed [5], of which Knowledge Distillation (KD) is one of the solutions [6]. The idea behind KD is similar to transfer learning. Not only can KD effectively reduce the size of the neural network and improve the inference efficiency, but it can also improve the accuracy in the case that training samples are insufficient and unbalanced, which may appear in DIL training phase. Hence, we believe that applying KD can benefit the deployment of DIL model.

In this article, we leverage the emerging edge computing paradigm and propose a framework based on DIL and KD which jointly considers available computation and communication resources and makes fine-grained offloading decisions for end devices. The objective of the proposed framework is to minimize the end-to-end latency of compute-intensive tasks on end devices. We use offloading decision instances to train our DIL model offline and compress the model to a lightweight one by KD for quickly making near-optimal offloading decisions online.

The rest of article is organized as follows. We briefly review related works in Section 2. We explain how to build a DIL model and use it in computation offloading decisions in Section 3. Then we describe how to use KD to further optimize the performance of the DIL model in Section 4. Numerical experiment results are shown in Section 5. At last we discuss some future directions and conclude in Section 6.

## 2 Related Work

### 2.1 Computation Offloading Strategies

To achieve lower latency or energy, mobile end devices usually choose to offload tasks to the cloud or edge servers. However, due to the complexity of network conditions in practice, for different devices at different times, the optimal computation offloading decisions are different. It is difficult to find this optimal decision in real time. Traditional computation offloading strategies are mostly based on mathematical modeling. Researchers in [7] study computation offloading problem in multi-user MEC environment. They firstly prove that finding the best offloading strategies in multi-channel and multi-user condition is NP-hard. Then they model this problem as an offloading game and design a distributed approach to reach the Nash equilibrium. Authors in [8] study offloading video objects detection tasks to cloud server. In [8], a big YOLO is deployed in cloud while a lite YOLO is deployed at end devices. Many factors such as bit rate, resolution and bandwidth are considered and the offloading problem is formulated into a multi-label classification problem. A near-optimal solution is found by an iteration approach and it successfully achieve higher accuracy in video objects detection. The main disadvantage of mathematical modeling methods is that their complexity is high, which may cause non-negligible inference delays and makes them not conducive to deploy in MEC network.

One of the most common compute-intensive tasks are DNN inference. On this type of task, many researchers study specialized computation offloading strategies. Kang et al. [14] proposes Neurosurgeon for DNN offloading. Neurosurgeon divides DNN into two parts. One part runs at end devices and the other runs at the cloud. This method reduces the calculation at end devices, trying to find a balanced point between computation and transmission. Neurosurgeon evaluates the latency of each DNN layer by regression models offline, and uses these models to calculate the best divided point online tailored to end devices' performance and bandwidth.

Recently, some researchers introduce DRL to find computation offloading strategies. In this case, the latency or energy consumption is served as agents' reward. Authors in [10] consider a condition of vehicular networks based on software defined network and jointly optimize networking, caching, and computer resource by a double-dueling deep-Q-network. The main drawback of DRL-based approaches in computation offloading is that the offline training and online inference takes much overhead. To tackle this challenge, we propose to utilize DIL for computation offloading, whose training cost and inference latency are significantly lower than DRL.

### 2.2 Deep Imitation Learning and Knowledge Distillation

Deep imitation learning (DIL) refers to training agents to imitate human's behaviors by a number of demos. Compared to DRL, training and inference time of DIL is much shorter. Authors in [13] build an edge computation offloading framework based on DIL. However, since DIL is based on DNN, if the size of DNN grows too large, it may still result in high inference delay. On this issue, we use Knowledge Distillation to compress the DIL model.

Knowledge Distillation (KD) is firstly proposed in [6], where the authors show that small DNNs can achieve approximately high accuracy as large DNNs with relatively less inference latency. This motivates us to compress the models to reduce inference delay with tiny accuracy loss. In KD, a large DNN is trained on a large training set and a lite DNN is trained on a small training set whose labels are the output of large DNN after "softened".

In our work, we compress our DIL model through KD to further reduce the inference delay, and improve model's performance when training samples are not enough or unbalanced.

## 3 Edge Computation Offloading by Deep Imitation Learning

### 3.1 System Model

We study the problem of making fine-grained offloading decisions for a single end device user. A compute-intensive task A on end device needs to be executed. We firstly spilt task A into some subtasks, follow [9]. Each subtask can be denoted by a tuple $a_t = (t, \varepsilon_t, d_t, d_{t+1})$. Task A can be seen as a set of all subtasks $a_t$. $\varepsilon_t$ represents the computation complexity of $t^{th}$ subtask (usually in CPU cycles). All the computation complexity forms a set $E = \{\varepsilon_t | t \epsilon [0, |A|)\}$. $d_t$ denotes the size of input data of $t^{th}$ subtask (usually in bytes). When t=0, $d_0$ represents the size of input data of task A. $d_{t+1}$ denotes the size of output data of $t^{th}$ subtask, and is also the input data size of $t + 1^{th}$ subtask. When t=|A|, $d_{|A|}$ represents output size data of task A. Sizes of all data flow jointly form the set $D = \{d_t | t \in [0, |A| + 1)\}$.

As is shown in Fig. 1, during the runtime of the mobile end device, it will establish a wireless connection with an edge server, and the edge server maintains a connection with the cloud server through the Internet. When a computation task in end device needs to be executed, it will be divided into some subtasks. Each subtask can choose to be executed locally on end device or sent to the edge server. When the edge server receives a requirement of execution of a subtask, it can decide whether to execute it locally on edge server or further send it to cloud server. Execution of a subtask leads to computation latency, which depends on the profile of end device and edge server and the computation complexity of subtasks E. If two adjacent subtasks are offloaded to different locations, transmission latency will also occur, which mainly depends on the bandwidth between end device, edge server and cloud server and transmission data size D. In this paper, considering the strong computing capability of the cloud server, cloud computation latency is far less than the transmission latency. Hence, when the subtask is offloaded to cloud server, the computation latency can be ignored and only the transmission latency is concerned.

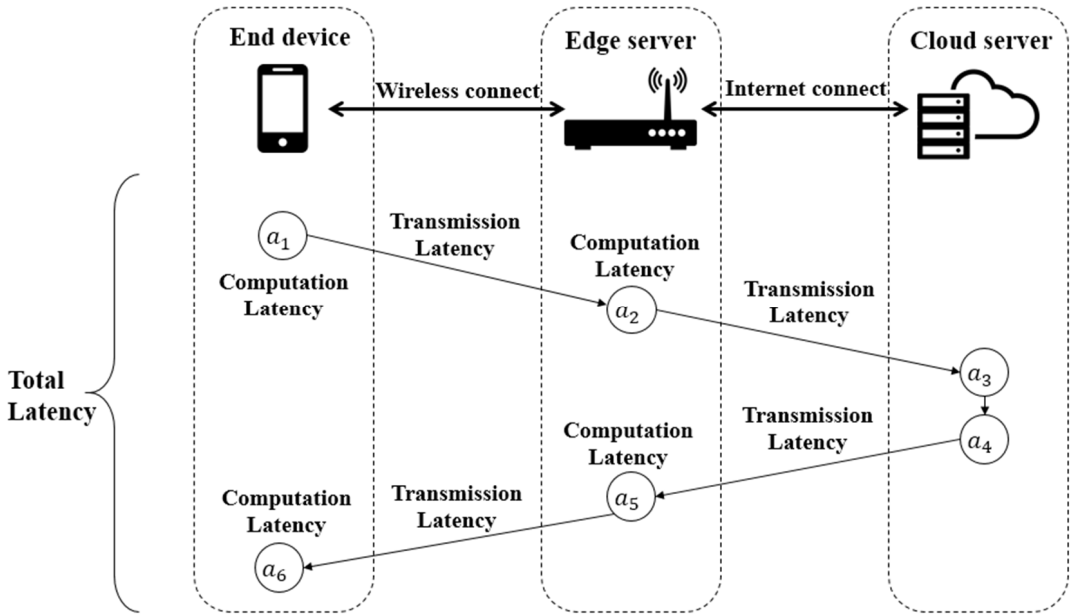

Fig. 1 Subtasks are offloaded to end device, edge server and cloud server respectively

### 3.2 Problem Formulation

When a computation task needs to be executed, end device split it into some subtasks and evaluate computation complexity E and transmission data sizes D of all subtasks. We can leverage the method introduced in [9] to evaluate E and D. Then all subtasks, E, D and the computing capability of end device (denoted by $p_1$) are sent to edge server. $p_1$ can be measured in CPU frequency (Hz). Edge server measures the bandwidth between end device and edge server (denoted by $b_1$) and bandwidth between edge server and cloud server (denoted by $b_2$). Factors mentioned above and the computing capability of edge server (denoted by $p_2$) jointly form the description of current offloading requirement $S = (E, D, p_1, p_2, b_1, b_2)$. Edge server are responsible to make offloading decisions of each subtask according to S.

For each subtask $a_t$, its offloading decision is represented by $I_t \in \{0,1,2\}$. $I_t = 0, 1, 2$ indicates that subtask $a_t$ is executed at end device, edge server or cloud server respectively. Offloading decision of the whole task A is given by $I = \{I_t | t \in [0, |A|)\}$. Obviously, $|I| = 3^{|A|}$. The offloading problem turns into finding the offloading decision I with the shortest end-to-end latency according to given S.

Now we compute the end-to-end latency of a specific I. As we have discussed, end-to-end latency can be divided into computation latency and transmission latency. Let $L_{exec}^t$ denote the computation latency of $t^{th}$ subtask. When $I_t = 0, 1$, subtask is executed at end device or edge server, hence $L_{exec}^t = \varepsilon_t/p_1$ or $L_{exec}^t = \varepsilon_t/p_2$, respectively. When $I_t = 2$, as is mentioned in Section 3.1, computation latency at cloud server is ignored, hence $L_{exec}^t = 0$. Given S and offloading decision I, computation latency of the whole task A is:

$$L_{exec}(S, I) = \sum_{t=0}^{|A|-1} L_{exec}^t.$$

Let $L_{trans}^t$ represent the data flow size between $t^{th}$ and $t - 1^{th}$ subtask. When data is transmitted between end device and edge server, $L_{trans}^t = d_t/b_1$ and when data is transmitted between edge server and cloud server, $L_{trans}^t = d_t/b_2$. Note that the data at the beginning of the whole task is input by the end device, and the final output destination is also the end device, we can assume that $I_{-1}$ and $I_{|A|}$ are always be 0. Given S and offloading decision I, transmission latency of the whole task A is:

$$L_{trans}(S, I) = \sum_{t=0}^{|A|} L_{trans}^t.$$

Our goal is to find the offloading decision $I^*$ with the shortest end-to-end latency, which is:

$$I^* = argmin_I(L_{exec}(S, I) + L_{trans}(S, I)).$$

So far, we have formulated computation offloading problem to an end-to-end latency minimization problem. By changing the parameter of argmin to energy, we can switch optimization objective to the energy consumption. Let S represent the description of offloading requirement, I represent the offloading decision, $R_{exec}(S, I)$ be the energy consumption of computation and $R_{trans}(S, I)$ be the energy consumption of transmission. Then the best offloading decision $I^*$ is: $I^* = argmin_I(R_{exec}(S, I) + R_{trans}(S, I))$. If it is required to optimize latency and energy simultaneously, we can set the parameter of argmin to a weighted sum of latency and energy.

### 3.3 Deep Imitation Learning for Offloading

The above minimization problem can be considered as a combinatorial optimization problem. Existing technologies such as traditional offloading algorithms or reinforcement learning are difficult to solve such problems efficiently. Hence, we first apply deep imitation learning (DIL) to deal with it based on the framework from [13]. Finding the best offloading decision $I^*$ can be formulated to a multi-label classification problem [11]. Decision I is a set of |A| labels and the three values of $I_t$ corresponding to three classes. The idea of DIL is to use a deep neural network (DNN) to learn the mapping from S to the best offloading decision $I^*$. To this end, offloading requirement S can serve as features of input samples and $I^*$ serves as the real labels of samples. As shown in Fig. 2.

DIL for offloading consists of three phases described follow:

1. Generate training samples offline: DIL is supervised learning and it needs a number of features labels pair $(S, I^*)$. The feature S can be obtained by collecting the actual offloading task requirement, or randomly generating features based on the distribution of various parameters in the actual offloading task requirement. Since labels $I^*$ are generated in an offline manner, some expensive non-real-time algorithm can be applied. In addition, performance of our DIL model is limited by the quality of labels, only the labels with high accuracy can ensure highly accurate DIL model. Note that the size of decision space is $3^{|A|}$. In summary, when |A| is small, we can use an exhaustive approach to obtain the optimal offloading decision by searching the whole decision space. When |A| is large, we solve this problem as integer programming problem by existing efficient solvers such as CPLEX.

2. Train DIL model offline: We build a DNN to learn the mapping from S to $I^*$. In this multi-label classification problem, the output of DNN consists of predictions of |A| labels. Each prediction has three possibility corresponding to three values of $I_t$. Hence the output layer of DNN has $3 \times |A|$ neurons and the activation function is SoftMax. All hidden layers are full connected layer.

3. Make offloading decisions online: After our DIL model is trained it is deployed to edge server to make offloading decisions online. Experiment shows that the efficiency of DIL model inference is higher than baseline models.

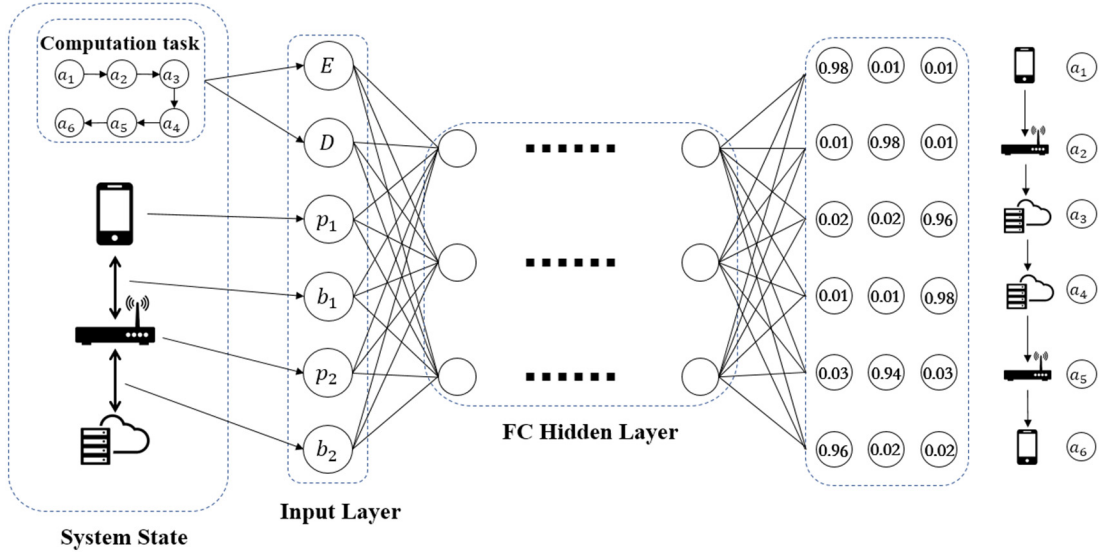

Fig. 2 Deep imitation learning model for edge computing offloading

DIL is based on learning. DIL's performance is closely related to the training samples. If the training samples are diverse, DIL model can deal with more conditions, i.e., more robust. If training samples contain offloading requirement under the conditions with fluctuation of wireless channels, DIL model can learn how to make a good decision under these conditions. In practice, training samples are from actual offloading requirement. The fluctuation of the wireless channels is also covered.

After the DIL model being trained, we should consider where the DIL model is deployed for online inference. Same as the computation tasks, DIL model can be deployed on end devices, edge server or cloud server. However, if DIL model is deployed on the cloud server, the wide-area connection will become an unstable factor. To ensure model's performance, we expect that the inference result of DIL model can be obtained with a low and predictable delay. Hence, even though the computing capability of cloud server is much stronger, it is not recommended to deploy DIL model on cloud server. In addition, since remaining all model inference workload on end device may lead to high energy consumption, we believe that edge server is a better place for DIL model deployment.

## 4 Knowledge Distillation for Model Compression

Since our DIL model is based on compute-intensive DNN execution, the inference latency could be high due to the limited computing capability of edge servers [15]. We hope that the DIL model running on the edge server is lightweight and the model inference delay is minimized [16]. Towards that, a potential solution is to put three phases mentioned above into edge server to train a DIL model based on small DNN locally on edge server. However, it raises two problems. First, with the limitation of number of parameters, the learning capability of a small DNN is insufficient. Compared to large DNN, it may cause loss of accuracy and make performance worse. Second, in the phase of generating demo offline, training samples are obtained by collecting the actual offloading task requirement or randomly generating based on distribution of various parameters in the actual offloading task. However, the service area of an edge server is highly limited. Compared to the samples collected by cloud server, samples collected by edge server may be not enough and unbalanced. This further incurs the accuracy and performance of small DNN. To this end, directly training a lightweight DIL model on edge server is not practical.

Authors in [6] proposed knowledge distillation (KD), which can be used for DNN compression. This technology helps us transfer the knowledge from a large DNN to a small DNN. When the training samples is not enough and unbalanced, accuracy of the DNN trained by KD is higher than the DNN directly trained on samples. Large DNN is called "teacher" and small DNN is called "student". Back to our offloading problem, we can leverage the strong computing capability of cloud server and a large number of samples to train a large DNN with high accuracy serve as teacher, and then transfer the knowledge learned by large DNN to small DNN which is deployed to edge server by KD, achieving low inference delay and small scale with little loss of accuracy, as shown in Fig. 3.

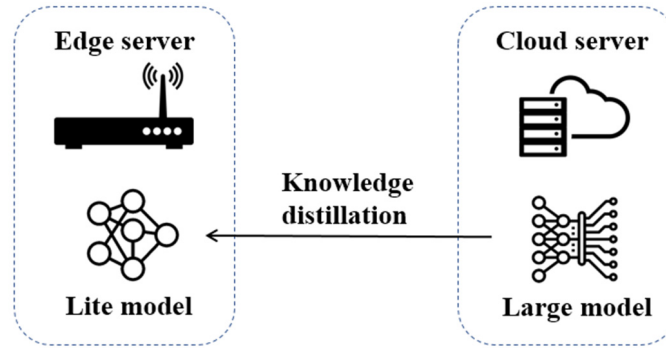

Fig. 3 Compress model by knowledge distillation to get a lightweight model deploying to edge server

KD can be applied to any neural networks whose output layer is activated by SoftMax, in other words, the networks used for solving classification problem. In KD, we train two networks, namely the teacher network and the student network. The teacher network is trained in a conventional way, while the student network is trained with the knowledge from the teacher. Specifically, before training the student network, we initialize the labels with the teacher network's output, rather than one-hot label from the training dataset.

In some cases, teacher network's output may be very small and close to zero (e.g., $< 10^{-3}$), which is nearly the same as the origin one-hot labels and remains difficulty for student network to learn the differences between labels. To alleviate this problem, we amplify the differences by further "softening" the labels. Let $p_i$ be the probability of $i^{th}$ class predicted by teacher, $q_i$ is the softened probability corresponding to $p_i$. We slightly change the form of the soften formula in [6] to compute $q_i$:

$$q_i = \frac{\exp(\ln(p_i)/T)}{\sum_{j=1}^{C} exp(\ln(p_j)/T)},$$

where C is the total number of classes (in our offloading problem C=3) and T is a tunable hyper-parameter with the constraint $T \geq 1$. If T=1, $q_i = p_i$. The labels will be softer with higher T. For instance, if original label is $(0.999, 2 \times 10^{-4}, 3 \times 10^{-6})$, when T=5, the soften label will be (0.71, 0.20, 0.09). When T=10, the soften label will be (0.53, 0.28, 0.19). In followed experiment we set T=5. Back to the offloading problem, we use a teacher network trained at cloud server to predict labels of the training set obtained by edge server. Then soften these labels by the formula mentioned above and train student network by softened labels at edge server.

We show the complete flowchart of our DIL offloading framework with KD in Fig. 4.

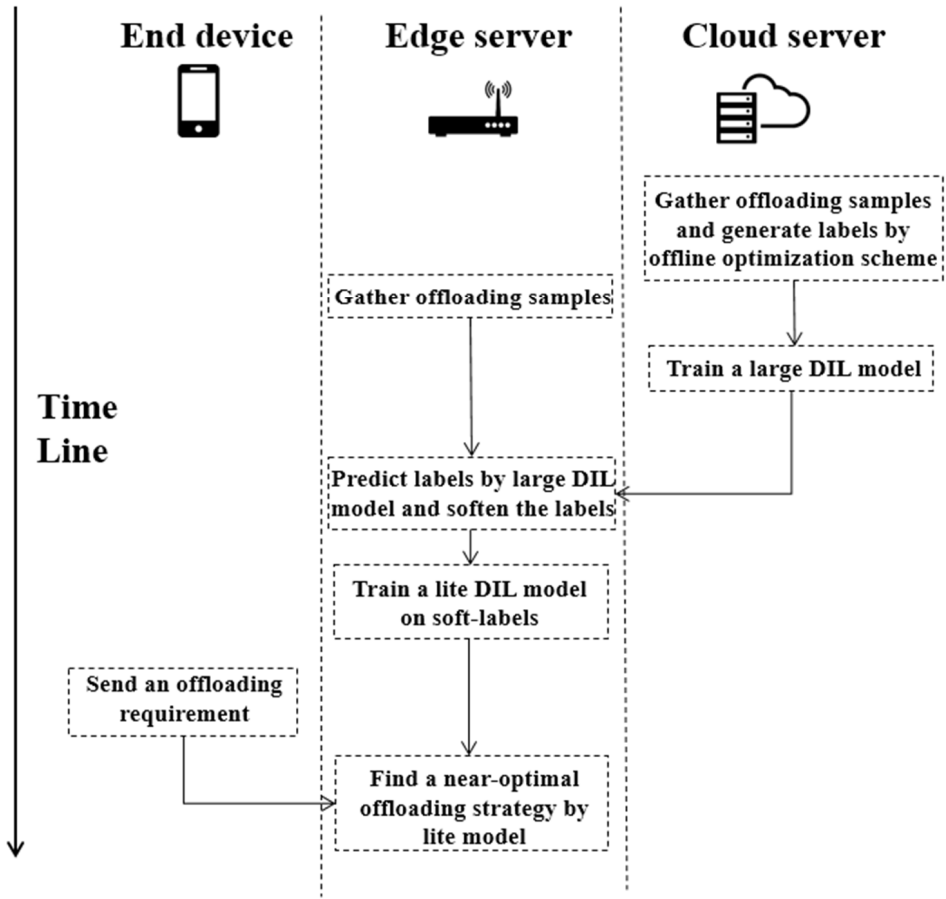

Fig. 4 Complete flowchart of our edge offloading framework based on DIL and KD

## 5 Evaluation

### 5.1 Evaluate Large DIL Model Performance

In this section, we set up numerical experiment to evaluate the performance of DIL model described in Section 3. We consider an MEC network consists of an end device user and an edge server connected by wireless connection, meanwhile the edge server connects to cloud server via the Internet. We assume that the compute-intensive task A on end device is divided into 6 subtasks, which is |A|=6. If the number of subtasks of some computation tasks is not 6, then we can merge some subtasks or insert empty subtasks to make the number of subtasks 6. The computation complexity of each subtasks $\varepsilon_t$ (measured in CPU cycles) are in the interval of $[0, 2000] \times 10^6$, following uniform distribution. Sizes of data transmission between subtasks follow uniform distribution with $d_t \in [0,10]$MB, like the setting in [12]. In addition, we assume that the computing capability of end device and edge server (both measured by CPU frequency in Hz) are in the intervals of $[100,1000]$MHZ and $[500,5000]$MHZ respectively, both following the uniform distribution. The bandwidth between end device and edge server and the bandwidth between edge server and cloud server are uniformly distributed in $b_1 \in [0,2]MB/s$ and $b_2 \in [0,3]MB/s$ respectively. We randomly generate 100K samples offline to train DIL model and 10K testing samples for testing.

Our DIL model is based on a DNN with 5 hidden layers. All hidden layers are fully connected layer and consist of 256 neurons. Number of parameters in the whole DNN is 1.6M. Activation function of hidden layers is RELU and output layer is activated by SoftMax. To evaluate the performance of our DIL based offloading framework, we consider some baseline frameworks listed below:

1. Optimal: Exhaustive method. For each sample, search the whole $3^{|A|}$ decision space, compute the latency described in Section 3.2 and choose the offloading decision with minimal latency. Note that this minimal latency is the lower bound in the decision space. Hence, this decision is bond to optimal.
2. Greedy: For each sample, find the offloading location one by one for each subtask to minimize the computation and transmission latency of current subtask.
3. DRL: Offloading framework based on deep reinforcement learning. Features of samples serve as environment and offloading decisions serve as actions. Opposite number of latency acts as reward. The deep Q network is similar to that in [10].
4. Others: Local: The whole task is executed on end device, which is for any t, $I_t = 0$. Edge: All subtasks are executed on edge server, which means $I_t = 1$. Cloud: All subtasks are offloaded to cloud server, which is $I_t = 2$. Random: Randomly choose offloading location for each subtask, that is to say $I_t$ are randomly choose from {0, 1, 2}.

Fig. 5 shows the normalized latency of DIL models and baseline frameworks with the latency of optimal decision is normalized to 1.0, then the latency of decision made by our DIL model is 1.095, with an increase less than 10%. Experiment results show that our model outperforms other baseline frameworks. Note that latency of "Edge" is less than "Local" and "Cloud", which indicates that edge server can certainly improve the compute-intensive tasks in end-to-end latency. At last, latency of "Random" is far higher than others, this is because randomly choosing offloading location will cause high transmission latency, which is expectable.

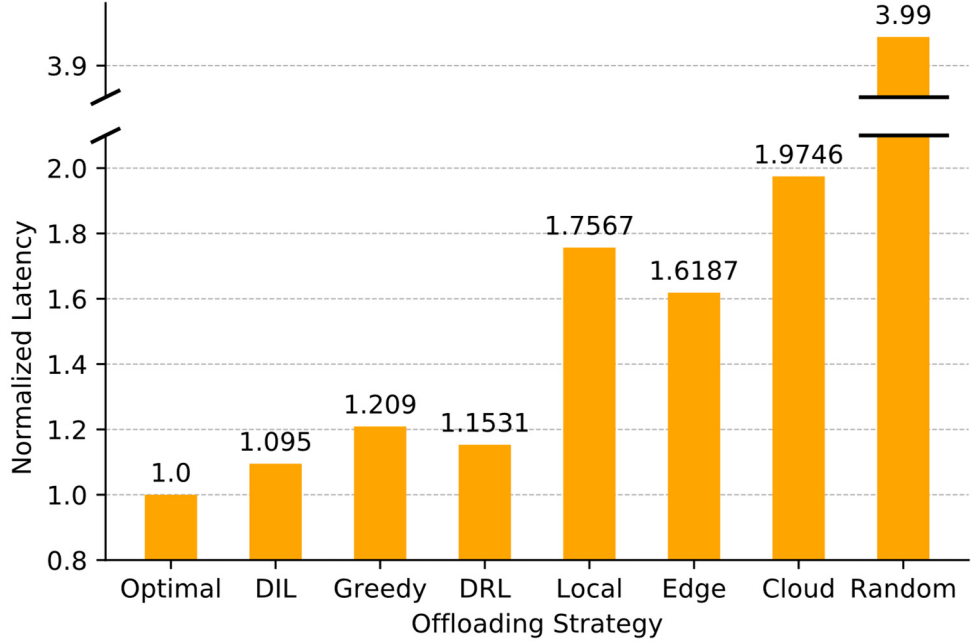

Fig. 5 Normalized end-to-end latency of offloading decisions made by our DIL model and baselines

**5.2 Evaluate Knowledge Distillation Performance**

As is mentioned in Section 4, we should compress our DIL model before deploying it to edge server and deal with the situation in which training samples on edge server are insufficient and unbalanced. We call our compressed model "KD-DIL" for short. In this section, we assume the CPU cycles of subtasks are uniformly distributed in $\varepsilon_t \in [500, 1500] \times 10^6$. Sizes of transmission data between subtasks is in $d_t \in [3, 8]MB$, following uniform distribution. The distribution range of $\varepsilon_t$ and $d_t$ are reduced by half compared to that in Section 5.1. Distributions of other parameters remain same. In order to simulate the case in which training samples are insufficient, we only generate 1K samples for training in this section, reduced by 99% compared to that in Section 5.1. Testing samples remain same as that in Section 5.1.

Our KD-DIL model is still based on DNN consisting of full connect layers. There are only 2 hidden layers in DNN with 32 neurons in each layer. The number of parameters of the whole DNN is about 10K, reduced by 99.375% compared to that in Section 5.1. Following baseline models are used for evaluating the performance of our KD-DIL model:

1. Baseline DIL: This DIL model is based on the DNN which is same as that in KD-DIL. The difference is that Baseline DIL is directly trained on the training set described above without applying KD described in Section 4.
2. DRL: Deep reinforcement learning based on DQN. The difference between this and DRL model in Section 5.1 is that it is trained on training set with 1K samples described above instead of that with 100K samples described in Section 5.1.
3. Greedy: Same as Greedy in Section 5.1.

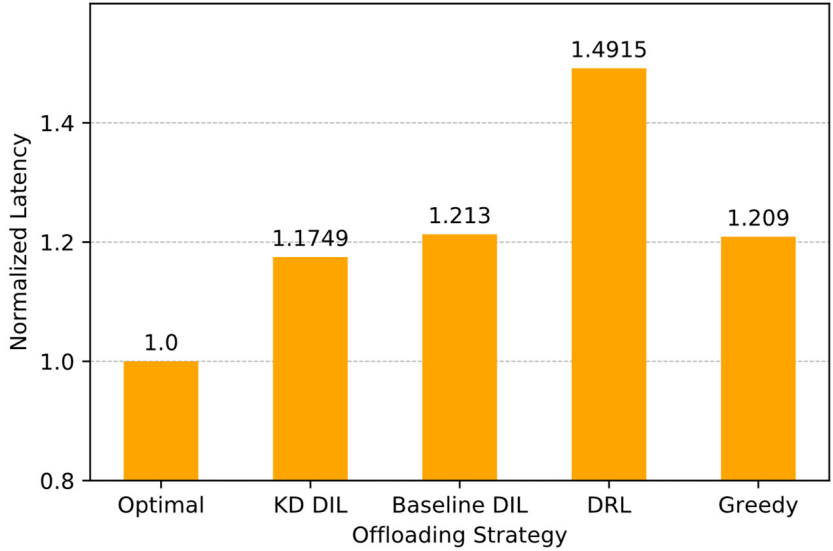

Fig. 6 Normalized KD-DIL model and baselines when using a small training set

Fig. 6 shows the normalized latency of KD-DIL models and baseline models. Again, the latency of optimal decision is normalized to 1.0. It shows that our KD-DIL model still outperforms baseline models. Not that the performance of DRL has a sharp decreasing compared that in Section 5.1 because of the change of training set. It is further shown that when number and distribution of training samples are changed, the accuracy loss of our KD-DIL model is relatively small.

At last, Table 1 shows the normalized inference delay of all models with delay of "Greedy" being normalized to 1.00, since greedy method is the most common method for computation offloading. We measured the delay of making 100K decisions of all models, and divide this delay by 100K to get the average delay of each decision. As shown in Table 1, compared to large DIL model, the inference delay of KD-DIL model decrease by 63% (0.17/0.51). Table 1 shows that the inference delay of Greedy approach is slightly higher than DIL model. As described in Section 5.1, Greedy approach finds deployment place for each subtask by iterations. The number of iterations equals to the number of subtasks. In practice, the number of subtasks may be much higher than 6, so the inference delay of Greedy approach may become much higher.

Lastly, the inference of optimal and DRL is hundreds of times that of our DIL models. Because optimal apply exhaustive method, high inference delay is expectable. While making decisions by DRL, we treat each strategy as an action and end-to-end latency as reward. We calculate each action's reward to find the highest reward, which needs many times of DNN inference. Hence, the delay of DRL inference is much higher than DIL.

Table 1: Inference delay of all models

| Model Name | KD-DIL | Large DIL | DRL | Greedy | Optimal |
| --- | --- | --- | --- | --- | --- |
| Normalized Delay | 0.17 | 0.51 | 119.72 | 1.00 | 122.54 |

## 6 Future Work and Conclusion

Flowcharts of subtasks can be represented by Directed Acyclic Graph (DAG) known as computation graph. In computation graph, nodes denote subtasks, edges denote data flow and directions of edge represent data transmission directions. DNN can also be regarded as a computation graph. In many programming frameworks dedicated for deep learning, e.g., TensorFlow, the concept of computation graph is applied. Offloading a computation graph in MEC network to optimize end-to-end latency is a difficult problem. The subtasks flowchart studied in this article has a list structure. In our future work we will focus on how to modify our work to adapt to DAG.

In this article, we have studied fine-grained edge computing offloading framework. In the situation, in which an end device wirelessly connects to an edge server, compute-intensive tasks can choose to be executed at end device, edge server or cloud server. We first review existing edge offloading framework including mathematic model method (game theory) and reinforcement learning. Then we provide model of computing task and describe the execution process of a task. Offloading problem is formulated into a multi-label classification problem and is solved by a deep imitation learning model. Next, in order to deal with the problem of training sample being insufficient and unbalanced, we apply knowledge distillation to get a lightweight model with little accuracy loss, making it easier to deploy to edge server. Numerical experiment shows that the offloading decisions made by our model have lowest end-to-end latency and the inference delay of our model is shortest, and after knowledge distillation we successfully reduce the inference delay by 63% with little accuracy loss. At last we briefly discuss some future directions of edge computation offloading.